# Correlation between metal-insulator transitions and structural distortions in high-electron-density SrTiO$_3$ quantum wells

Jack Y. Zhang[1], Clayton A. Jackson[1], Ru Chen[2], Santosh Raghavan[1], Pouya Moetakef[1),a)], Leon Balents[3], and Susanne Stemmer[1]

[1] Materials Department, University of California, Santa Barbara, California 93106-5050, USA

[2] Department of Physics, University of California, Santa Barbara, California 93106, USA

[3] Kavli Institute of Theoretical Physics, University of California, Santa Barbara, Santa Barbara, California 93106, USA

[a] Present address: Department of Chemistry and Biochemistry, University of Maryland, College Park, MD, USA

**Abstract**

The electrical and structural characteristics of $SmTiO_3/SrTiO_3/SmTiO_3$ and $GdTiO_3/SrTiO_3/GdTiO_3$ heterostructures are compared. Both types of structures contain narrow $SrTiO_3$ quantum wells, which accommodate a confined, high-density electron gas. As shown previously [Phys. Rev. B **86**, 201102(R) (2012)] $SrTiO_3$ quantum wells embedded in $GdTiO_3$ show a metal-to-insulator transition when their thickness is reduced so that they contain only two SrO layers. In contrast, quantum wells embedded in $SmTiO_3$ remain metallic down to a single SrO layer thickness. Symmetry-lowering structural distortions, measured by quantifying the Sr-column displacements, are present in the insulating quantum wells, but are either absent or very weak in all metallic quantum wells, independent of whether they are embedded in $SmTiO_3$ or in $GdTiO_3$. We discuss the role of orthorhombic distortions, orbital ordering, and strong electron correlations in the transition to the insulating state.

# I. Introduction

Quantum-confined transition metal oxides allow for creating new states of matter through manipulation of spin and orbital order, interfacial proximity effects, and reduced dimensionality, and can thus serve to elucidate the physics of two-dimensional, strongly-correlated electron systems [1]. For example, narrow, high electron-density quantum wells of a non-magnetic band-insulator, $SrTiO_3$, which are embedded in a Mott insulating ferrimagnet, $GdTiO_3$, show ferromagnetism and mass enhancement due to strong electron correlations [2-4]. At the smallest dimensions, when the quantum wells contain just two SrO layers, the electron system abruptly localizes and the resistivity increases by several orders of magnitude [2]. The transition to the insulating state is accompanied by structural distortions of the Ti-O octahedra, which can be experimentally detected by measuring concurrent displacements of the Sr cations [5]. Metal-insulator transitions at reduced thicknesses have also been observed in narrow quantum wells and thin films of many other perovskite materials, such as $SrVO_3$ [6], $LaNiO_3$ [7-9], and $NdNiO_3$ [10]. In general, in many $d$-electron systems, symmetry breaking of spin and orbital degrees of freedom play a crucial role in promoting an insulating state in materials that undergo a metal-insulator transition [11]. Transition metal-oxygen octahedral tilts that reduce the symmetry relative to the parent cubic perovskite structure are modified in quantum wells due to film strain [12,13] and interfacial coherency [3,14-16].

To understand the underlying physics of Mott transitions in confined quantum wells, such as the relative roles of disorder, the interactions among the electrons themselves (strong correlations), and interactions of the carriers with the lattice, it is useful to explore if the localization can be systematically tuned by changing the external parameters of the system. Towards this goal, we compare the electrical and structural properties of thin SrTiO$_3$ quantum wells embedded in GdTiO$_3$ and SmTiO$_3$, respectively. We have previously reported on the electrical and structural properties of the structures with GdTiO$_3$ [2,5], and they are included here for comparison. In both cases, the quantum wells contain a two-dimensional electron gas with sheet carrier densities of close to one electron per (pseudo-)cubic planar unit cell, which is introduced by the charge discontinuity at the interface [2,17]. This sheet carrier density is independent of the film thicknesses. It is important to emphasize that SrTiO$_3$ is a band insulator in bulk, and has the ideal cubic perovskite structure at room temperature. Therefore, and in contrast to the aforementioned confined correlated metals, such as the nickelates, correlated properties, including magnetism, mass enhancement and metal-insulator transitions, are induced in a material that does not exhibit Mott physics in the bulk. Both GdTiO$_3$ and SmTiO$_3$ are prototypical Mott insulators, with a $d^1$ electron configuration. SmTiO$_3$ has the same orthorhombic crystal structure as GdTiO$_3$, albeit with slightly smaller octahedral distortions [18]. The two compounds also differ in their low-temperature magnetic properties - GdTiO$_3$ is ferrimagnetic while SmTiO$_3$ is antiferromagnetic [19]. These properties couple with the electron system in the quantum well [4]. Furthermore, they exhibit different orbital ordering, which is antiferro-orbital in GdTiO$_3$ and ferro-orbital in SmTiO$_3$, respectively [20-22].

## II. Experimental

All films were grown by hybrid molecular beam epitaxy (MBE) [23,24] on (001) (La$_{0.3}$Sr$_{0.7}$)(Al$_{0.65}$Ta$_{0.35}$)O$_3$ (LSAT, a=3.86 Å) substrates. Electrical measurements were carried out on GdTiO$_3$/SrTiO$_3$/GdTiO$_3$ and SmTiO$_3$/SrTiO$_3$/SmTiO$_3$ quantum well structures that contained a single SrTiO$_3$ quantum well. The GdTiO$_3$ and SmTiO$_3$ layers were 10 nm thick. The thicknesses of the SrTiO$_3$ quantum wells are specified in terms of the number of SrO planes, as verified by transmission electron microscopy (TEM). Electrical contacts, consisting of a 40 nm Ti and 400 nm Au top layer, were deposited by electron beam evaporation in van der Pauw geometry. A Physical Property Measurement System (Quantum Design PPMS Dynacool) was used for resistivity and Hall measurements.

For the TEM studies, multilayer structures were grown to allow for the characterization of all SrTiO$_3$ quantum well thicknesses with the same TEM sample [see Figs. 1(a-b)]. TEM cross-sections were prepared by focused ion beam thinning with 5 kV Ga ions and imaged using a field emission FEI Titan S/TEM operating at 300 kV with a super-twin lens ($C_s = 1.2$ mm). For high-angle annular dark-field (HAADF) imaging in scanning TEM (STEM) a convergence semi-angle of 9.6 mrad was used. HAADF-STEM images were taken at the same magnification, with a frame size of 1024 × 1024 pixels, and a dwell time of 30 μs. A-site cation displacements (where A represents Gd, Sm or Sr in the chemical formula ATiO$_3$), which directly correlate with the octahedral tilts [18,25-27], were characterized by measuring the deviation angle, 180°-$\theta$, where $\theta$ is the angle formed between three successive A-site cations, averaged over multiple HAADF images [5]. SmTiO$_3$ and GdTiO$_3$ films were oriented such that the longest axis of the orthorhombic unit cell (*c*-axis in the P*bnm* space group notation) was in the plane of the film [24]. The average in-plane strain of coherent films in this orientation is approximately -0.6% and -1% for GdTiO$_3$ and SmTiO$_3$, respectively.

Because films were grown on a cubic substrate, they contained four symmetry-related orientation variants [24]. Images for analysis were taken along [110]$_o$, where the subscript indicates the orthorhombic unit cell. While MBE offers atomic layer control, quantum well width variations of ±1 atomic planes are unavoidable, due to surface steps and substrate miscut along the projection of the sample. Only regions with layer thicknesses corresponding to the nominal thickness were chosen for analysis.

Experimental deviation angles were compared with results from density functional theory (DFT) calculations of periodic superlattices containing two SrO layers embedded in four layers of SmTiO$_3$, (SrTiO$_3$)$_2$(SmTiO$_3$)$_4$. DFT calculations were performed in the Wien2k [28] implementation and the generalized gradient approximation (GGA) [29]. The calculations used a $2a \times 2a \times c$ unit cells to allow for octahedral tilts, and $a$ was set to the experimental LSAT lattice constant, 3.86 Å. Structure optimization was done both on the atomic coordinates and the $c/a$ ratio, within the GGA+$U$ approximation, as described in detail elsewhere [3]. We applied $U_{eff} = U - J = 3.5$ eV on the Ti $d$ orbitals and $U_{eff} = 8.5$ eV on the Gd and Sm $f$ orbitals. Atomic relaxations on the superlattice were performed until the Hellmann-Feynman forces on atoms were less than 5 meV/Å. We note that the calculations were carried out for the experimentally observed orientation relationships. This is in contrast to the calculations in ref. [3], in which the orthorhombic $c$-axis (P$bnm$ space group notation) was perpendicular to the quantum well plane.

## III. Results and Discussion

Figures 1(a-b) show HAADF-STEM overview images and schematics of the multilayer structures used for measuring the deviation angles. The GdTiO$_3$ and SmTiO$_3$ layers were 4 nm thick, while the SrTiO$_3$ quantum well thickness was varied from one to eight SrO layers (in the sample with GdTiO$_3$) and two to five SrO layers (in the sample with SmTiO$_3$). Both structures had 10 nm SrTiO$_3$ buffers and caps, respectively.

Figure 2 shows the measured deviation angles, integrated across each atomic plane, taken from regions of the samples such as those indicated by the boxes Fig. 1(a). Both plots contain data averaged over multiple images from different regions of each sample to reduce noise and improve sampling. The corresponding HAADF intensity profiles (intensity averaged over a five pixel radius around each centroid position [30], square symbols) are plotted above the deviation angles. The pronounced atomic-number contrast of HAADF allows for identification of the layers from the intensities. The SrO planes, identified by their lower HAADF intensity, are highlighted in Fig. 2. The deviation angles are larger in the center of GdTiO$_3$ and SmTiO$_3$ layers, and closely match those expected from the bulk values for a coherently strained film in each case (15.7° and 14.7° for GdTiO$_3$ and SmTiO$_3$, respectively) [27]. The deviation angles in the interior of the SmTiO$_3$ layers are smaller than those in GdTiO$_3$, as expected from their respective bulk structures. The SrO layers show no deviations from the cubic structure (deviation angle ≈ 0°) for all quantum well thicknesses greater than two SrO layers, as reported previously for SrTiO$_3$ quantum wells in GdTiO$_3$ [5,27]. The apparent deviation angle of ~1.5° is due to a combination of scan noise and experimental instabilities, as it is measured even in the buffer and capping layers, as well as in unstrained SrTiO$_3$ (not shown). As reported previously, for a two-SrO-thick quantum well in GdTiO$_3$, significant Sr-site displacements are observed, indicative of octahedral distortions and an orthorhombic-like structure [5]. In contrast, the two-SrO quantum well in SmTiO$_3$ shows only a very slight increase in the deviation angle, indicating that the Ti-O-Ti bond angles remain close to the 180° angle in cubic SrTiO$_3$.

We briefly discuss why interfacial intermixing or roughness cannot be responsible for the measured structural distortions. Gd$_x$Sr$_{1-x}$TiO$_3$ alloys remain cubic up to Gd concentrations of $x = 0.3$ [31]. Therefore very large concentrations of Gd intermixing would need to be present to induce an orthorhombic distortion. HAADF-STEM contrast is highly sensitive to the atomic number, and while the contrast is dependent on the dopant position along the beam direction [32,33], a Gd concentration of 30%, given the TEM sample thicknesses used here (~15-20 nm), would be easily detectable from image intensities [34]. In Fig. 1(c) the square symbols show image intensities normalized for each sample. The two-SrO-layer quantum wells show similar intensities as the five-SrO-layer quantum wells (dashed boxes), indicating similar chemical composition. The intensity in the center of the five-SrO-layer quantum well serves as a reference for pure SrTiO$_3$ intensity, as it agrees with intensities in the buffer and capping layers (not shown), after accounting for the TEM sample thickness. The data point marked by an arrow in Fig. 2 indicates an intermixed atomic layer, which was discernible by eye in the HAADF image. The intermixed layer also has a smaller deviation angle. The intensities from the 2 SrO layer are *lower* than this intermixed layer, yet show much *higher* deviation angles, indicating that the distortion is not an effect of disorder or intermixing.

The deviation angles calculated from DFT agree well with the experimental results. Shown in Fig. 3 are DFT results for the deviations angles for samples with quantum wells containing two SrO layers and comparisons with the experimental results. The DFT calculations showed that the Ti-O-Ti bonds (not shown) are less distorted in the two SrO layers embedded in the SmTiO$_3$, which results in the smaller deviation angles, as seen in Fig. 3. The calculations underestimate the distortions in the quantum wells in GdTiO$_3$. DFT also slightly underestimates the orthorhombic distortions in bulk GdTiO$_3$ and slightly overestimates the orthorhombic distortion in bulk SmTiO$_3$ relative to the experimental values [18] (the deviation between DFT calculations and experiments in Ti-O-Ti bond angles is less than 2°).

Figure 4 shows the sheet resistances as a function of SrTiO$_3$ quantum well thickness and temperature. All structures contain a high-density electron gas (carrier density ~ $7 \times 10^{14}$ cm$^{-2}$), as a consequence of the interface doping, which resides entirely within the SrTiO$_3$ [2]. Metallic behavior is observed for all quantum wells in SmTiO$_3$, down to the thinnest limit of a single SrO layer. In contrast, quantum wells embedded in GdTiO$_3$ are metallic for thicknesses greater than two SrO layers, but become insulating at lower thicknesses. Comparing Figs. 2 and 4, we see that the metal-insulator transition that occurs in the thinnest GdTiO$_3$-embedded quantum wells directly correlates with the presence of the structural distortion, (relatively) large octahedral tilts and reduced Ti-O-Ti bond angles, while the metallic behavior in SmTiO$_3$-embedded quantum wells over all thickness ranges correlates with a (relative) lack of structural distortion and close to 180° bond angles.

Although disorder (i.e. chemical mixing at the interface, SrTiO$_3$ thickness fluctuations) likely exists in both types of samples, and indeed plays a role in low-temperature transport [2], the results shown in Figs. 2 and 4 provide evidence that the metal-insulator transition is caused by true Mott-Hubbard-like physics. Specifically the *abrupt transition to an insulator* as the thickness is changed by a single atomic plane is associated with an *abrupt structural transition* that cannot have been caused by disorder.

The results also offer additional insights into the correlation physics of the perovskite titanates. Large octahedral tilts and reduced Ti-O-Ti bond angles, which directly correlate with the transition to the insulating state, occur only in the quantum wells embedded in $GdTiO_3$, despite similar electron densities in quantum wells in both types of structures. It is already known from bulk materials that small differences in octahedral distortions around a critical value are associated with large effects on the transport properties. Distortions serve to decrease the Ti-O-Ti bond angles, lifting the $t_{2g}$ orbital degeneracy and reducing the Ti $3d$ bandwidth. A crossover from large to small polaron transport occurs in the lightly-doped, insulating rare earth titanates between $SmTiO_3$ and $GdTiO_3$ [19]. We note, however, that all bulk rare earth titanates are insulating at all temperatures. The electrons in the insulating quantum wells of $SrTiO_3$ in $GdTiO_3$ form a high-density small polaron gas [35]. In contrast, the much smaller distortions in the quantum wells in $SmTiO_3$ are correlated with an electron gas that never self-localizes.

While the results provide evidence for the crucial role of the Ti-O-Ti bond angles in the insulating state, the underlying origins of the quantitative differences in the degree of structural distortion appears to be more complex. Ti-O-Ti bond angles are only slightly smaller in $GdTiO_3$ (145.76° basal, 143.87° apical) than in $SmTiO_3$ (147.29° basal, 146.48°) [18]. To maintain coherent bonding at the interface, these distortions may couple to the Ti-O-Ti bond angles in the $SrTiO_3$ quantum well. However, as previously noted, interfacial coupling seems to be mostly accommodated via reduced distortions in the rare earth titanate, at least for sufficiently thick SrTiO$_3$ [27]. Furthermore, we note that the deviation angle difference between the two bulk structures (~1°) is significantly smaller than the deviation angle difference between the two-SrO-layer thick quantum wells in the two structures (~3°). This suggests that there are additional factors at play that promote the larger distortions in the thinnest quantum wells in GdTiO$_3$, which, incidentally, are also not completely captured by DFT.

One possibility is strong correlations that may also drive orbital order, which then couples with the structure. In the antiferromagnetic rare earth titanates, Takubo et al. have reported a crossover from antiferro-orbital to ferro-orbital order at temperatures significantly above the Néel temperature [20]. It is therefore possible that coupling with the antiferro-orbital ordering favors a more distorted state in the quantum wells in GdTiO$_3$. While the crossover temperature, a prime example of the strong interplay between spin and orbital fluctuations with the perovskite lattice, occurs below room temperature for bulk SmTiO$_3$, it could be shifted to higher temperatures by coherency strain that exist in the thin film structures. We note, however, that the question whether orbital ordering occurs above the magnetic ordering temperatures is still under debate [36]. The results support a view of electron-electron interactions in the quantum wells driving the structural distortions at least to a certain degree.

## IV. Conclusions

In summary, we have shown that a metal-insulator transition that is observed below a critical thickness in high-electron-density SrTiO$_3$ quantum wells occurs only if orthorhombic-like structural distortions are sufficiently large. The degree of distortion is controlled by the specific rare earth titanate that interfaces the quantum well. Using STEM, we showed that the metal-to-insulator transition is an intrinsic phenomenon that is correlated with a symmetry lowering structural distortion, indicative of Mott-Hubbard-like behavior, with disorder playing (at most) a secondary role. Specifically, even a single SrO layer embedded in SmTiO$_3$ remains metallic. The degree of the observed distortion is larger than what would be expected from simple lattice geometrical considerations based on the bulk structure, and suggests that more complex physics such as strong electron correlations and orbital order influences the structure in the quantum well. Future studies should investigate the influence of interface orientation, since this may result in different coupling between the octahedral tilts in the quantum well and the Mott insulator.

**Acknowledgements**

The microscopy experiments were supported by DOE (Grant No. DEFG02-02ER45994). Film growth, transport experiments and theory were supported by a MURI program of the Army Research Office (Grant No. W911-NF-09-1-0398). J.Y.Z. received support from the Department of Defense through an NDSEG fellowship, and C.A.J. from the National Science Foundation through a Graduate Research Fellowship. This work made use of facilities from the UCSB Materials Research Laboratory, an NSF-funded

MRSEC (DMR-1121053), as well as the UCSB Nanofabrication Facility, a part of the NSF-funded NNIN network.

**Figure Captions**

**Figure 1:** (Color online) (a) HAADF-STEM images and (b) schematics of multilayer structures with SrTiO$_3$ quantum wells embedded in GdTiO$_3$ and SmTiO$_3$ layers. The brighter regions in (a) are GdTiO$_3$ or SmTiO$_3$ layers, while darker regions are SrTiO$_3$ layers. The labels in (b) indicate the thicknesses of the SrTiO$_3$ layers, measured in number of SrO planes. The GdTiO$_3$ and SmTiO$_3$ layers were 4 nm thick.

**Figure 2:** (Color online) Deviation angles (red circles) in each AO plane and corresponding normalized HAADF intensities (blue squares), for regions containing two SrO and five SrO layers [indicated by the white boxes in Fig. 1(a)]. SrO layers are highlighted in gold. The dashed lines serves as guides to mark structural distortions (or lack of) in the SrTiO$_3$ wells. The dashed boxes indicate atomic planes of similar intensity. The data for the structures with GdTiO$_3$ was previously shown in ref. [5].

**Figure 3:** (Color online) Comparison between experimental (circles) and calculated (DFT, squares) deviation angles for 2 SrO layers (shaded) between (a) GdTiO$_3$ and (b) SmTiO$_3$.

**Figure 4:** (Color online) Temperature dependent sheet resistance for GdTiO$_3$/SrTiO$_3$/GdTiO$_3$ (top) and SmTiO$_3$/SrTiO$_3$/SmTiO$_3$ (bottom) structures as a function of SrTiO$_3$ layer thickness. A metal-insulator transition occurs in GdTiO$_3$/SrTiO$_3$/GdTiO$_3$ structures when the SrTiO$_3$ thickness is reduced to two SrO layers. The data for the structures with GdTiO$_3$ was previously shown in ref. [2].

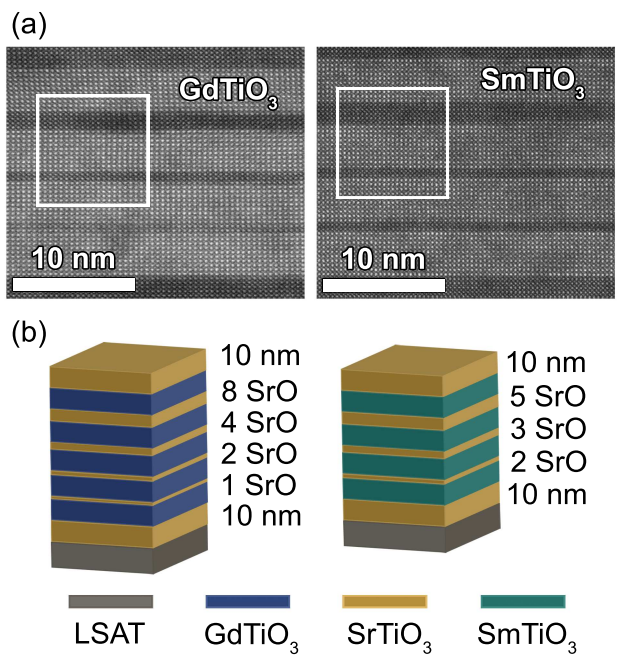

Figure 1

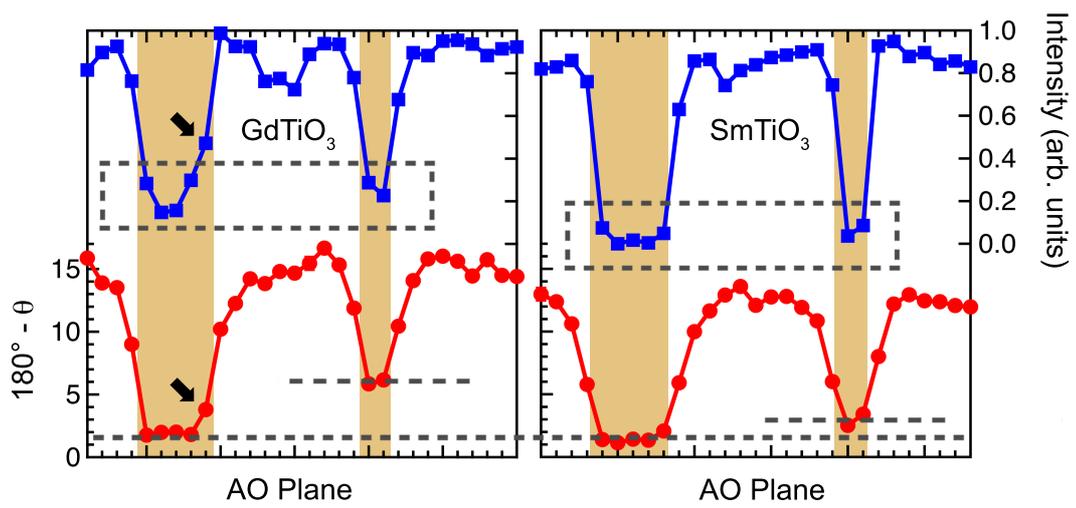

Figure 2

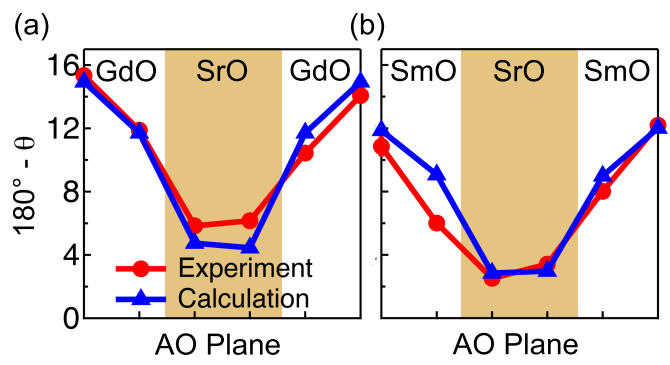

Figure 3

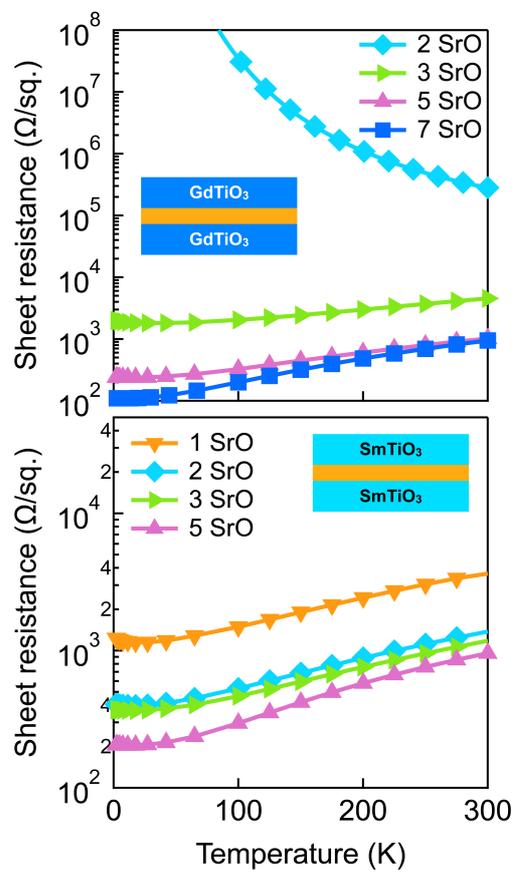

Figure 4